\documentstyle[12pt]{article}

\newcommand{\be}{\begin{equation}}
\newcommand{\ee}{\end{equation}}
\newcommand{\ba}{\begin{eqnarray}}
\newcommand{\ea}{\end{eqnarray}}
\newcommand{\spz}{\hspace{0.7cm}}
\newcommand{\virg}{\spz,\spz}
\newcommand{\pvirg}{\spz;\spz}

\newcommand{\la}{\lambda}
\newcommand{\dx}{\partial_x}
\newcommand{\dt}{\partial_t}
\newcommand{\dla}{\partial_{\lambda}}

\newcommand{\lt}{\left(}
\newcommand{\rt}{\right)}

\newcommand{\bC}{\mbox{\bf C}}

\newcommand{\bZ}{\mbox{\bf Z}}
\newcommand{\buno}{\mbox{\bf 1}}

\newcommand{\cL}{{\cal L}}

\newcommand{\NP}[1]{Nucl.\ Phys.\ {\bf #1}}
\newcommand{\PL}[1]{Phys.\ Lett.\ {\bf #1}}

\newcommand{\CMP}[1]{Comm.\ Math.\ Phys.\ {\bf #1}}

\newcommand{\PRL}[1]{Phys.\ Rev.\ Lett.\ {\bf #1}}

\newcommand{\IJMP}[1]{Int.\ J.\ Mod.\ Phys.\ {\bf #1}}

\newcommand{\LMP}[1]{Lett.\ Math.\ Phys.\ {\bf #1}}

\begin{document}
\sloppy
\renewcommand{\thefootnote}{\fnsymbol{footnote}}

\newpage
\setcounter{page}{1}

\vspace{0.7cm}
\begin{flushright}
DTP-98-67\\
September 1998
\end{flushright}
\vspace*{1cm}
\begin{center}
{\bf Non-Local Virasoro Symmetries in the mKdV Hierarchy}
        \footnote{Work supported  by European Union under contract
FMRX-CT96-0012} \\
\vspace{1.8cm}
{\large D.\ Fioravanti $^a$ and M.\ Stanishkov $^b$ \footnote{On leave
from I.N.R.N.E. - Sofia, Bulgaria\\ 
E-mail: Davide.Fioravanti@durham.ac.uk, marian@lpm.univ-montp2.fr}}\\
\vspace{.5cm}
$^a${\em Department of Mathematical Sciences - Univ. of Durham\\
     South Road, DH1 3LE DURHAM, England} \\
$^b${\em Laboratoire de Physique Mathematique - Univ. de Montpellier II\\
     Pl. E. Bataillon, 34095 Montpellier, France} \\
\end{center}
\vspace{1cm}

\renewcommand{\thefootnote}{\arabic{footnote}}
\setcounter{footnote}{0}

\begin{abstract}
{\noindent We generalize} the dressing symmetry construction in mKdV
hierarchy. This leads to non-local vector fields (expressed in terms of
vertex operators) closing a Virasoro algebra. We argue that this algebra
realization should play an important role in the study of 2D integrable
field theories and in particular should be related to the Deformed
Virasoro Algebra (DVA) when the construction  is perturbed out of the
critical theory.
\end{abstract}
\newpage

The search for underlying symmetries is a powerful strategy for the
investigation of Quantum Field Theories (QFT's). The most popular example
is probably 2D Conformal Field Theory (CFT) \cite{BPZ}. Its covariance
under the infinite dimensional Virasoro symmetry leads sometimes to its
complete solution.
The situation is more involved for more general Integrable Field 
Theories (IFT). They are characterized by an infinite number of local
integrals of motion in involution. However, this infinite symmetry does
not carry sufficient information because it is abelian. Nevertheless most
of these theories possess a bigger non-local and nonabelian symmetry like
the Yangian symmetry of the Heisenberg spin chain \cite{B} or
$\hat{sl}_q(2)_{k=0}$ in the sine-Gordon theory \cite{BL}. 
Usually these symmetries relate the correlation functions of fields in the
same multiplet without giving any explicit information on the properties
of these correlators \cite{LS}.   
In addition, another interesting symmetry, connected with the elliptic
deformation of Virasoro algebra (DVA) \cite{SKAO}, is suspected to play an
important role in various theories \cite{Lu}. At present a lot is known
about 
the mathematical structure of DVA, for instance HW representations,
intertwiners (or vertex operators) etc. . What remains unclear is its
realisation in the  physical  picture.

Two different possible treatments of IFT were suggested recently. In a
series of papers Bazhanov, Lukyanov and Zamolodchikov \cite{BLZ} proposed
an approach to the description of CFT which is a variant of the inverse
scattering method. The main idea is to express the basic objects at the
classical level (KdV theory) in terms of the scattering data encoded in
the transfer matrix $T(x;\lambda)$. The quantization prescription is to
replace the underlying $A_1^{(1)}$-loop algebra by its q-deformation and
the mKdV field  $\phi(x)$ by the quantum free scalar field. The
off-critical theory is described by the {\it transfer matrix} ${\bf
T}(x,\bar x;\lambda,\mu)=
\bar T(\bar x;\mu/\lambda)T(x;\lambda)$ where $\bar T$ is the antichiral
transfer matrix and $\mu$ is the coupling constant.

Another powerful approach to the study of the classical integrable systems
is based on the so called dressing transformations \cite{Sem,BB}. These
are essentially gauge symmetries of the related linear problem and lead to
Poisson-Lie type transformations of the basic fields of the theory. It was
also suggested \cite{BB} that (almost) all the solutions of the
corresponding equations of motion lie in the orbits of the dressing group.

In a recent article we proposed an approach to the study of IFT which is
an attempt of unifying the aforementioned approaches. We showed, on the
example of (generalized) (m)KdV systems,
that the dressing by means of the regular transfer matrix leads to a
symmetry which has a Poisson-Lie action on the  basic field and so is the
ancestor of the quantum group symmetry at the quantum level. The
asymptotic dressing instead leads naturally to a new description of the
{\it spectrum} of the (m)KdV theory. Indeed, in the case of
$A^{(1)}_1$-KdV this spectrum is exactly the classical limit of the
construction \cite{BBS}. A number of constraints (or classical
null-vectors) appears in this construction. In our approach they appear as
constraints or equations of motion for the transfer matrix and the
resolvent 
$Z(x;\lambda)$. We proposed further to quantize and extend out of the
critical point this construction following the mentioned prescription
\cite{BLZ}.

In this letter we propose a
generalization of the aformentioned dressing transformations. The central
idea is that one can dress not only the generators of the underlying
Kac-Moody algebra but also a differential operator in the spectral
parameter: 
$\lambda^m\dla^n$. The corresponding vector fields close a $w_\infty$
algebra with a Virasoro subalgebra for the case $n=1$. The general lines
of the construction were described earlier \cite{OS}. Here we specialize
it to the specific case of the mKdV system and show the arising of a set
of new non-local vector fields from the regular dressing. They are
expressed in terms of vertex operators and complete the asymptotic ones,
which were known \cite{D}, to the full Virasoro algebra. This is important
in view of the possibility of allowing a central extension.

We suggest that this Virasoro symmetry could be of great importance for
the 
study of IFT. First of all it provides a new set of conserved charges
closing a nonabelian algebra, thus carrying nessesarily more information
about the theory. Furthermore one can quantize it (i.e. find its
realization in the conctest of CFT) and then perturb it out of criticality
by following the BLZ prescription \cite{BLZ}. We claim that the perturbed
version should be closely related to the aforementioned DVA. In fact, one
can consider the Virasoro algebra we are going to construct as a field
theory version of that which appeared in the study of Calogero-Sutherland
model. Actually, its deformation in this model led just to the discovery
of the DVA \cite{SKAO}.

Let us briefly recall the construction of \cite{FS1} for the case of mKdV
equation. This non-linear equation can be written as a zero-curvature
condition:
\be
[ \dt - A_t , \dx - A_x ] = 0 
\label{lax}
\ee
for connections belonging to the $A_1^{(1)}$  loop algebra 
\ba
A_x &=& - v h + (e_0 + e_1),  \nonumber \\
A_t &=& \la^2(e_0 + e_1 - vh) -
\frac{1}{2}[(v^2-v')e_0 + (v^2+v')e_1] - \frac{1}{2}(\frac{v''}{2}-v^3)h
\label{con}
\ea
where the generators $e_0,e_1,h$  are chosen in the
fundamental representation and canonical gradation 
of the $A_1^{(1)}$ loop algebra
\be
e_0=\la E=\left(\begin{array}{cc} 0 & \la \\
                                  0 & 0 \end{array}\right)\virg
e_1=\la F=\left(\begin{array}{cc} 0 & 0 \\
                                  \la & 0 \end{array}\right)\virg
h  = H=\left(\begin{array}{cc} 1 & 0 \\
                               0 & -1 \end{array}\right) .
\label{gen}
\ee
The KdV variable  $u(x,t)$ is related to the mKdV variable
$v(x,t)$  by the Miura transformation \cite{Miu} 
$u=-v^2 +v'$.  

A remarkable geometrical interest is obviously attached to 
the transfer matrix which performs the parallel transport 
along the $x$-axis, and is thus the solution of the associated linear
problem
\be
\dx T(x;\la) = A_x(x;\la)T(x;\la)  . 
\label{lap}
\ee
The solutions of the previous equation are holomorphic functions of
$\lambda\in{\bC}$ and exhibit an essential singularity at
$\lambda=\infty$. Consequently they admit two different kinds of
expansion: the first as a convergent series in positive powers of
$\lambda$ with an infinite  
radius of convergence and the second as asymptotic expansion around
$\lambda=\infty$. In this article we will stress briefly how these two
kinds of expansion furnish a unified procedure to derive symmetries of
classical non linear systems written in Lax form (\ref{lax}) completing
the constuction of \cite{FS1}. A longer exposition of these topics will be
given in \cite{35}.

A solution of (\ref{lap}) of the first kind can be written as 
\be
T(x,\la) =e^{H\phi(x)}{\cal P}
\exp\lt\la \int_0^xdy
(e^{-2\phi(y)} E+ e^{2\phi(y)} F ) \rt =\left(\begin{array}{cc} A & B \\
                                                                C & D
\end{array}\right) ,
\label{firsol}
\ee
where $A=e^{\phi}(1+\sum_1^{\infty}\lambda^{2n}A_{2n}),
B=e^{\phi}\sum_0^{\infty}\lambda^{2n+1}B_{2n+1}, C[\phi]=B[-\phi],
D[\phi]=A[-\phi]$.
Let us construct the generic  resolvent by dressing one of the generators
$X=H,E,F$ (\ref{gen}) by mean of the regular expansion (\ref{firsol})
\be 
Z^X(x,\la)=(TXT^{-1})(x,\la)=\sum_{k=0}^\infty \la^k Z^X_k  . 
\label{dres}
\ee
The observation that $Z^X(x,\la)$ is clearly a  resolvent for the operator
${\cL}=\dx -A_x$   (\ref{con}), i.e. it satisfies, by construction, 
\be
[{\cL},Z^X(x;\la)]=0  ,
\label{resdef}
\ee
motivates the construction of candidates for gauge connections of the
dressing symmetries
\be
\Theta^X_{-n}(x;\la)=(\la^{-n}Z^X(x;\la))_-=\sum_{k=0}^{n-1} \la^{k-n}
Z^X_k  .
\label{theta}
\ee
In fact, this observation ensures that the r.h.s. of the following gauge
transformation is of degree zero in $\lambda$
\be
\delta^X_{-n} A_x = -[\Theta^X_{-n}(x;\la),{\cL}]. 
\label{gautra1}
\ee
This transformation will be consistent if the last term in (\ref{gautra1})
is
proportional to $H$: $\delta^X_{-n} A_x=H \delta^X_{-n}  \phi'$ .
This depends, for X fixed, on whether n is even or
odd \cite{FS1}:
\begin{itemize}
\item in the $\Theta^H_{-n}$ case $n$, in
(\ref{gautra1}), must be even,
\item in the  $\Theta^E_{-n}$ and  $\Theta^F_{-n}$ case $n$ must
conversely be odd. 
\end{itemize}
It is possible to show  by direct calculation  that these
infinitesimal transformation generators form a representation of  a
(twisted) Borel subalgebra of the loop algebra $A_1^{(1)}$
\be
[\delta^X_{-m} , \delta^Y_{-n}] = \delta^{[X,Y]}_{-m-n} \pvirg X,Y = H,E,F
.
\label{delalg}
\ee
The first generators of this algebra are explicitly given by: 
\ba
\delta^E_{-1} \phi'(x) &=&  e^{2\phi(x)}          \nonumber  \\
\delta^F_{-1} \phi'(x) &=&  - e^{-2\phi(x)}    \nonumber  \\
\delta^H_{-2} \phi'(x) &=& e^{2\phi(x)} \int_0^x dy e^{-2\phi(y)} +
e^{-2\phi(x)} \int_0^x dy 
e^{2\phi(y)}
\label{che}
\ea
and the rest are derived from these by commutation.
It was shown in \cite{FS1} that these transformations are not symplectic
but are rather generated via Poisson-Lie action of 
the coresponding non-local charges.

An asymptotic form solution of (\ref{lap}) plays instead a  crucial role 
in obtaining the local integrals of motion and the isospectral flows
\cite{Fa84,DS}. The asymptotic expansion for a solution of (\ref{lap}) can
be written as  \cite{FS1}  
\be 
T(x;\la)=KG(x;\la)e^{-\int_0^x dy D(y)},
\label{asyexp}
\ee
in terms of a constant matrix  
$
K =\frac{\sqrt{2}}{2}\left(\begin{array}{cc}  1 & 1 \\
                                              1 & -1 \end{array}\right)$,
of a diagonal matrix 
\be
D(x;\la)=d(x;\la)H , \spz
d(x;\la)=\sum_{k=-1}^{\infty}\la^{-k}d_k(x)  
\label{lod}
\ee
containing the local conserved densities $d_{2n}(x)$ and, finally, of the
off-diagonal matrices $G_j(x), j>0$ 
\be
G(x;\la)=\buno + \sum_{j=1}^{\infty}\la^{-j}G_j(x)
\ee
with entries $(G_j(x))_{12}=g_j(x)$ and $(G_j(x))_{21}=(-1)^{j+1}g_j(x)$.
It is likewise known \cite{DS} that the construction of the mKdV flows
goes
through the definition of a  resolvent $Z(x;\la)$ characterized by the
following property of its asymptotic expansion 
\be 
[{\cL},Z(x;\la)]=0,\spz
Z(x,\la)=\sum_{k=0}^\infty \la^{-k} Z_k, \spz Z_0=E+F ,  
\label{asyres} 
\ee
which implies the modes of the $\la$-expansion to have the form 
\be 
Z_{2k}(x)=b_{2k}(x)E+c_{2k}(x)F \virg Z_{2k+1}(x)=a_{2k+1}(x)H . 
\label{solasy} 
\ee 
Now it is possible to reinterpret the definition (\ref{asyres}) in terms
of
{\it dressing} through the asymptotic expansion of $T$ (\ref{asyexp}):  
\be
Z(x,\la)=(THT^{-1})(x,\la) . 
\label{asydre} 
\ee
As in the regular case, the system enjoys a gauge symmetry 
of the form (\ref{gautra1}) with $X=H$ and $n=2k+1$,
$\theta_{2k+1}$ are the Lax
connections associated to the mKdV-flows $\delta_{2k+1}$:
\be
\theta_{2k+1}(x;\la)=(\la^{2k+1}Z(x;\la))_+=\sum_{j=0}^{2k+1}
\la^{2k+1-j} Z_j(x) .  
\ee 
It turns 
out that this gauge transformation coincides 
with the commuting higher mKdV flows:  
\be 
\delta_{2k+1}\phi'(x)=\partial a_{2k+1}(x).
\label{mkdvh} 
\ee 
Finally, we proposed in \cite{FS1} an alternative description of the
spectrum of local fields. We proposed to use as basic objects the entries
of the resolvent $Z(x;\la)$ modulo the gauge transformations described
above. A number of constraints, or {\it classical null vectors}, appear in
this picture coming from the equation of motion
$\delta_{2k+1}Z=[\theta_{2k+1},Z]$ of the resolvent and the obvious
constraint $Z^2=\buno$.

At this point we are in a position to construct in a natural way more
general kinds of dressing-like symmetries. 
 It is well known that the vector fields $l_m=\lambda^{m+1}\dla$ on the
circumpherence realize the centerless Virasoro algebra 
\be
[l_m,l_n]=(m-n)l_{m+n}.
\ee 
A very natural dressing is represented by the resolvent
\be
Z^V_{-m}=T_{reg}l_{-m}T^{-1}_{reg} , \spz m>0
\label{rvr}
\ee 
where $T_{reg}$ indicates the regular expansion of the transfer matrix
(\ref{firsol}) and $m$ is a positive integer. Of course, this dressed
generator satisfies the usual property (\ref{resdef}) of being a resolvent
\be
[{\cL},Z^V_{-m}(x;\la)]=0 .
\label{resdef1}
\ee
As in the previous cases, this property (\ref{resdef1}) allows us to
calculate the expansion modes $Z_n$ of
\be
Z^V_{-1}=T_{reg}l_{-1}T^{-1}_{reg}=\sum_{n=0}^{\infty}\la^n Z_n -\dla
\ee
and thus the expansion modes of the more general Virasoro resolvent
(\ref{rvr}). In the same way, (\ref{resdef1}) authorizes us to define a
gauge connection 
\be
\theta^V_{-m}=(Z^V_{-m})_- =\sum_{n=0}^m\la^{n-1-m}Z_n -\dla
\ee
and the relative gauge transformation
\be
\delta_{-m}^V A_x =-[\theta^V_{-m}(x;\la),{\cL}] .
\label{rvf}
\ee
Finally, we have to verify the consistency of this gauge transformation
requiring $\delta_{-m}^V A_x=H \delta_{-m}^V  \phi'$ .
It is very easy to see that this requirement imposes $m$ to be even.
Explicit examples of the first flows are
\ba
\delta_{-2}^V \phi' &=&  e^{2\phi(x)} \int_0^x dy e^{-2\phi(y)} -
e^{-2\phi(x)} \int_0^x dy e^{2\phi(y)}=e^{2\phi(x)}B_1-e^{-2\phi(x)}C_1
\nonumber  \\ 
\delta_{-4}^V  \phi' &=&  e^{2\phi(x)}(3B_3(x)-A_2(x)B_1(x)) -
e^{-2\phi(x)}(3C_3(x)-D_2(x)C_1(x)) \nonumber  \\
\delta_{-6}^V  \phi' &=& e^{2\phi(x)} (5B_5(x)-3A_4(x)B_1(x)+A_2(x)B_3(x))
\nonumber  \\
&-& e^{-2\phi(x)}(5C_5(x)-3D_4(x)C_1(x)+D_2(x)C_3(x))  .  
\label{frd's} 
\ea
We stress that these infinitesimal variations have a form very similar to
that of the regular dressing flows ((\ref{gautra1}) with $X=H$)
$\delta_{-2r}^H$. Nevertheless, in spite of the commutativity
$[\delta_{-2r}^H,\delta_{-2s}^H]=0$ 
one can check by direct calculation that instead the flows (\ref{frd's})
obey Virasoro commutation relations:
$[\delta_{-2},\delta_{-4}]=\delta_{-6}$.
Actually, this is true also in the general case:
\be
[\delta_{-2m}^V ,\delta_{-2n}^V ]=(2n-2m)\delta_{-2m-2n}^V .
\label{rsa}
\ee
The validity of this formula can be proved using a generalization of the
proof scheme shown in \cite{DS} for the mKdV isospectral flows. We shall
return to this point in the longer article \cite{35}. 
From the actions (\ref{frd's}) the transformations of the classical {\it
primary fields} 
$e^{\phi}$ follow. For example:
\ba
\delta_{-2}^V e^{\phi} &=& (D_2-A_2)e^\phi  \nonumber  \\
\delta_{-4}^V e^{\phi} &=&  [(3D_4-C_3B_1)-(3A_4-B_3C_1)]e^\phi .
\nonumber  
\ea
It is understood of course that these fields are primary with respect to
the usual {\it space-time} Virasoro symmetry.

In the same way it is quite natural to generate a resolvent by dressing
the remaining vector fields $l_m=\lambda^{m+1}\dla$, $m\geq 0$
\be
Z^V_{m}=T_{asy}l_{m}T^{-1}_{asy} ,\spz   m\geq 0
\label{avr}
\ee 
through the asymptotic expansion of the transfer matrix (\ref{asyexp})
$T_{asy}$. Now we have:
\be
Z^V_{-1}=T_{asy}l_{-1}T^{-1}_{asy}=\sum_{n=0}^{\infty}\la^{-n} Z_n -\dla.
\ee
In general :
\be
Z_{2n}=\beta_{2n}E+\gamma_{2n}F ,\spz  Z_{2n+1}=\alpha_{2n+1}H \nonumber
\ee
Where for example $\beta_0=x=\gamma_0$, $\alpha_1=2xg_1$,
$\beta_2=-xb_2-g_1+\int^x d_1$, $\gamma_2=-xc_2+g_1+\int^x d_1$ etc. .
In the same manner we define a gauge connection 
\be
\theta^V_{m}=(Z^V_{m})_+=\sum_{n=0}^{m+1}\la^{m+1-n}Z_n -\dla
\ee
and the relative gauge transformation
$\delta_{m}^V A_x =-[\theta^V_{m}(x;\la),{\cL}]$.
The consistency condition of this gauge transformation,
$\delta_{m}^V A_x=H \delta_{m}^V  \phi'$,
impose $m$ to be even in this case too. Explicit examples of the first
flows are
\ba
\delta_{0}^V \phi' &=& \phi' + x\phi'' \nonumber \\
\delta_{2}^V  \phi' &=& 2xa_3'+6g_3-2g_1^3+2g'_1\int^x d_1 .   
\label{fad's} 
\ea
For the simplest primary field these imply:
\ba
\delta_{0}^V e^{\phi} &=& (x\dx+\Delta)e^\phi  \nonumber  \\
\delta_{2}^V e^{\phi} &=& (2xa_3+2g_2+2g_1\int^x d_1)e^\phi .  
\ea
One can easily find the {\it equation of motion} for the mKdV resolvent:
$\delta^V_{2m} Z=[\theta^V_{2m},Z]$. In particular:
\be
2\delta_2b_2=\delta_2 u= x(u'''-{3\over 2}uu')+u''-2u^2-{1\over
2}u'\int^xu . 
\ee
It coincides exactly with the already known expression \cite{GO}.

One can check by direct calculation that the first flows (\ref{fad's})
obey Virasoro commutation relations. Actually, in general one can show
that:
\ba
\delta_{2n}^VZ_{2m}^V&=& [\theta^V_{2n},Z^V_{2m}]-(2n-2m)Z^V_{2n+2m}
\nonumber \\
\delta_{2n}^V\theta_{2m}^V-\delta_{2m}^V\theta_{2n}^V&=&
[\theta^V_{2n},\theta^V_{2m}]
-(2n-2m)\theta^V_{2n+2m} .
\ea
From these it is not difficult to see that the asymptotic flows also close
(half) the Virasoro algebra:
\be
[\delta_{2m}^V ,\delta_{2n}^V ]=(2m-2n)\delta_{2m+2n}^V ,\spz m,n\ge 0.
\label{asa}
\ee 

An important question arises at this point: what are the commutation
relations between the asymptotic and regular transformations? This is a
very nontrivial question in view of the different character of the
corresponding vector fields - the asymptotic ones are quasilocal (they can
be made local by differentiating a certain number of times), the regular
instead are essentially non-local being expressed in terms of vertex
operators. We recall here that the (proper) regular dressing symmetries
(\ref{gautra1}) commute with all the mKdV flows (\ref{mkdvh}) \cite{FS1}.
We shall see that this is not the case here. In fact it is easy to compute
the most simple relations: $[\delta_0,\delta_{2n}]=-2n\delta_{2n}$, $n\in
{\bf Z}$ (i.e. $\delta_0$ counts the dimension or level). Using the
explicit formulae presented above one can also compute the first
nontrivial commutator: $[\delta_2,\delta_{-2}]=4\delta_0$. In fact, one
can show that in general \cite{35} :
\ba
\delta^V_mZ^V_{-n}&=& [\theta^V_m,Z^V_{-n}]-(m+n)Z^V_{m-n} \nonumber \\
\delta^V_{-m}Z^V_n&=& [\theta^V_{-m},Z^V_n]+(m+n)Z^V_{n-m}\nonumber  \\
\delta^V_m\theta^V_{-n}-\delta^V_{-n}\theta^V_m&=&
[\theta^V_m,\theta^V_{-n}]-(m+n)\theta^V_{m-n} 
\ea
From these it is easy to deduce:
$[\delta^V_m,\delta^V_{-n}]=(m+n)\delta^V_{m-n}$, and therefore:
\be
[\delta^V_{2m},\delta^V_{2n}]=(2m-2n)\delta^V_{2m+2n}, \spz m,n\in\bZ .
\label{vir}
\ee
We want to stress once more that this Virasoro symmetry is different from
the {\it space-time} one and is essentially non-local. The additional
symmetries coming from the regular dressing are very important for
applications. They complete the asymptotic ones forming an entire Virasoro
algebra and provide a possibility of a central extension. However, this
central term may appear only in the algebra of the generators of the above
transformations, a subject that we leave for a future publication
\cite{35}.

With the aim of understanding the classical and quantum structure of the
mKdV system we present here the complete algebra of symmetries. The
Virasoro flows commute neither with the mkdV hierarchy (\ref{mkdvh}) nor
with the (proper) regular dressing flows (\ref{gautra1}). In fact one can
show that \cite{35} :
\be
[\delta_{2k+1},\delta^V_{2m}]=(2k+1)\delta_{2k+1+2m} , \spz
[\delta^X_n,\delta^V_{2m}]= n \delta^X_{n+2m} .
\label{virdr} 
\ee
Note that the indices of r.h.s. can become negative, for the first, or
positive, for the second of equations (\ref{virdr}). Explicit calculation
show that in these cases the commutator is exactly equal to zero. This
fact confirms the self-consistensy of the construction.

One remark is necessary at this stage. One can consider the flows
constructed in this letter as a special case of true symmetries, commuting
with the flows of the mKdV hierarchy, with the times $t_{2n+1}, n\ge 1$
set to zero. In the case of the asymptotic flows one can restore the time
dependance by 
adding a series containing explicitly the mKdV times $t_{2k+1}$ 
\be
\delta^V_{2m} \rightarrow \delta^V_{2m} - \sum_1^{\infty}(2k+1)t_{2k+1}
\delta_{2m+2k+1},
\ee
the resulting Virasoro flows commute with the mKdV flows $\delta_{2k+1}$.

In conclusion we have presented a generalization of the dressing symmetry
construction leading to a non-local Virasoro symmetry of the mKdV
hierarchy. We stress that it has nothing to do with the space-time
Virasoro, generated at the classical level by the moments of the {\it
classical stress tensor} $\int x^n u(x)dx$. It is obtained instead by
dressing the differential operator $\lambda^{n+1}\partial_{\lambda}$. In
view of the relation between the spectral parameter and the on-shell
rapidity $\lambda= e^\theta$, it is generated probably by diffeomorphisms
in the momentum space and in this sense is dual to the space-time Virasoro
symmetry. We suggest that it is a remnant of the Virasoro symmetry present
in the  specific case of the 
massive Ising model as constructed in \cite{SSS}.
Although we presented a construction only in the case of mKdV it can be
easily extended for the generalized KdV theories as well. Of particular
interest is the $A^{(2)}_2$ hierarchy, connected with the $\phi_{1,2}$
perturbation of CFT models \cite{FRS}.

Actually, the asymptotic Virasoro symmetry was already known. It plays an
important role in the study of the matrix models where it leads to the so
called Virasoro constraints: $L_n\tau=0, n\ge 0$. Here $\tau$ is the
$\tau$-function of the hierarchy and is connected to the partition
function of the matrix model. Moreover, it seems that it should play an
important role also in the conctest of the Matrix String Theory \cite{YM},
which is now intensively studied.
Note also that these Virasoro constraints are the classical version of the
conditions for the HW state. It will be very interesting to investigate
the action of the rest of Virasoro charges $L_n, n<0$ on the
$\tau$-function.

Furthermore, such a symmetry appears also in the study of
Calogero-Sutherland model whose connection with the matrix models and CFT
is well known. Moreover, it is known that in the q-deformed case it
becomes a deformed Virasoro algebra \cite{SKAO}. It is natural to suppose
that in the same way our construction is deformed off-critically. Let us
note also that actually the symmetry  of mKdV is much larger. It is
obtained by dressing more general operators - 
$\lambda^m\partial_\lambda^n$, the corresponding vector fields closing a
$w_\infty$ algebra. In particular those coming from
$\lambda^n\partial_\lambda^n$ all commute thus providing an infinite
number of commuting (non-local) Hamiltonians. We suggest that they are
connected to the higher Calogero-Sutherland Hamiltonians in their
collective field theory description. 

Recently, Babelon, Bernard and Smirnov \cite{BBS} constructed certain
null-vectors off-criticality in the context of the form-factor approach.
They showed that there is a deep connection, at the classical level,
between their construction and the finite zone solutions of KdV and the
Witham theory of evaraged KdV. On the other hand the Virasoro algebra
presented above has a natural action on the finite zone solutions,
changing the complex structures of the corresponding hyperelliptic Riemann
surfaces, and on the basic objects of the Witham hierarchy \cite{GO}. This
suggests that exists a deep connection between this Virasoro symmetry and
the null-vectors of \cite{BBS}. 

In this connection we note the work \cite{BB} on the creation of N-soliton
solutions of KdV out of the vaccum by means of dressing transformations.
What is the action of our Virasoro on the vaccum and the N-solitons? This
is an important question since the form-factors can be viewed as a
quantization of the N-soliton solutions \cite{BBS2}. 
We shall try to answer all these questions in a forthcoming paper
\cite{35}.

{\bf Acknowledgments} - We are indebted to E. Corrigan, Al. Zamolodchikov
and V. Fateev
for discussions and interest in this work. D.F. thanks 
the EC Commission for financial support through TMR Contract
ERBFMRXCT960012. M.S. thanks the Department of Mathematical Sciences of
Durham for the warm hospitality during which this work has been completed.

\end{document}